\documentclass[12pt]{article}

\usepackage{amsmath,amssymb}
\usepackage{graphics,epsfig}
\usepackage{curves}

\newcommand{\E}{{\boldsymbol{E}}}
\newcommand{\X}{{\boldsymbol{X}}}
\newcommand{\Y}{{\boldsymbol{Y}}}
\newcommand{\el}{{\boldsymbol{\ell}}}
\newcommand{\n}{{\boldsymbol{n}}}
\newcommand{\m}{{\boldsymbol{m}}}
\newcommand{\q}{{\bar{\boldsymbol{m}}}}
\newcommand{\hp}{{\hat{\Psi}}{}}

\begin{document}
\noindent
gr-qc/0012088 \hfill Class. Quantum Grav. {\bf 18}, 3059 (2001)

\begin{center}
{\Large {\bf Classification of image distortions in terms 
\\[0.2cm]
of Petrov types}}
\\[0.4cm]
{\large {\em Thoralf Chrobok}}
\\
TU Berlin, Sekr.\ PN 7-1, 10623 Berlin, Germany.
\\
tchrobok@itp.physik.tu-berlin.de
\\[0.4cm]
{\large {\em Volker Perlick}}
\\
Albert Einstein Institute, 14476 Golm, Germany.
\\
(Permanent address: TU Berlin, Sekr. PN 7-1, 10623 Berlin, Germany.
vper0433@itp.physik.tu-berlin.de)
\end{center}


\begin{abstract}
    An observer surrounded by sufficiently small spherical light
    sources at a fixed distance will see a pattern of elliptical
    images distributed over the sky, owing to the distortion effect
    (shearing effect) of the spacetime geometry upon light bundles.
    In lowest non-trivial order with respect to the distance, this
    pattern is completely determined by the conformal curvature
    tensor (Weyl tensor) at the observation event. In this paper
    we derive formulae that allow one to calculate these distortion
    patterns in terms of the Newman-Penrose formalism. Then we
    represent the distortion patterns graphically for all Petrov
    types, and we discuss their dependence on the velocity of the
    observer.


\medskip\noindent
        Suggested short title: Classification of image distortions

\end{abstract}


\section{Introduction}\label{sec:intro}
The general-relativistic light deflection, i.e., the influence
of the spacetime geometry on the paths of light rays, has the effect
that, in general, the apparent shape of a distant object at the
celestial sphere of an observer will be distorted. This distortion
effect is directly related to the {\em shear\/} of light
bundles, whereas the apparent size and the apparent brightness
of images are related to the {\em expansion\/} of light bundles. By the
well-known {\em Sachs equations} \cite{Sa}, the shear is an effect
of the conformal curvature tensor (Weyl tensor), whereas the expansion
is influenced by the Ricci tensor along the light rays. In the
following we concentrate on the distortion effect, i.e. on the shear.
In a pioneering paper, Kristian and Sachs \cite{KS} introduced a certain
measure for the distortion effect, given in terms of a series expansion
with respect to the distance between source and observer, and suggested
using it as a cosmological observable. In lowest non-trivial order with
respect to the distance the distortion is completely determined by
the Weyl tensor at the observation event. It is then a natural question
to ask to what extent the distortion is already determined by
the Petrov type of the Weyl tensor. Up to now only some special aspects
of this question have been addressed in the literature (see Penrose and
Rindler \cite{PR}, Volume II, Chapter 8); it is the purpose of this
paper to give a more detailed discussion, including graphical
representations of the distortion patterns on the sky for each
Petrov type.

To gain understanding the distortion effect it is didactically helpful to
begin the discussion with an over-idealized universe in which all
galaxies are perfectly spherical. We shall assume that all light rays
that reach the observer from a particular galaxy can be viewed as
`infinitesimally close to each other', i.e., that the laws of
light propagation can be linearized around a central light ray.
Then the image of each galaxy results by applying a linear map to
a circle, i.e., it is an ellipse. By the term `distortion pattern'
we mean the distribution of those ellipses over the celestial sphere
of the observer, for galaxies at a specific distance. (We are interested
only in the shape, not in the size, of the ellipses.) Having fixed the
distance, the distortion pattern is given by a non-negative scalar function
on the celestial sphere that gives the eccentricity of the respective
ellipse, and by a direction field on the celestial sphere that
indicates the major axis of the respective ellipse. The relevant
formulae for determining these fields, which are due to Kristian and
Sachs \cite{KS}, will be given in Section \ref{sec:derivation} below.
We shall restrict our discussion to the case that only terms of
lowest non-trivial order with respect to the distance are to be taken
into account. (In this approximation it does not matter which of
the various non-equivalent general-relativistic notions of `distance'
we use, see Section~\ref{sec:derivation} below.)
The distortion pattern is then completely
determined by the Weyl tensor at the observation event, with the
eccentricities depending quadratically on the
distance and the direction field being independent of the distance.
If the Weyl tensor is zero, then there is no distortion effect, i.e.,
the eccentricity is everywhere zero and the direction field is
undetermined. If the Weyl tensor is not zero, then the distortion
effect vanishes at exactly four (not necessarily distinct) points on
the sky, corresponding to the four principal null directions of the
Weyl tensor.
Kristian and Sachs \cite{KS} concentrate on the magnitude of the
distortion effect, i.e., on the eccentricities which
they measure in terms of a function they denote by $e$. Penrose and
Rindler \cite{PR}, Volume II, Chapter 8, on the other hand, concentrate
on the direction field which they call the `fingerprint'
of the Weyl tensor. It is our goal to discuss both the magnitude
and the direction of the distortion effect; therefore, we assign
to each point of the observer's celestial sphere a `distortion length
element' whose length gives the eccentricity and whose direction gives
the major axis of the respective ellipse. In Section~\ref{sec:derivation}
we show how these quantities can be calculated with the help of the
Newman-Penrose formalism. In Section~\ref{sec:discussion}
we discuss, in terms of pictures, the `distortion length element field'
and its dependence on the oberserver's velocity for each Petrov type.

If we want to link these deliberations with the real universe we have
to face the problem that galaxies are not spherical. We may restrict
to those galaxies which appear elliptical on the sky, to within
reasonable approximation, and we may assume that the actual shape of
those galaxies is approximately that of a rotational ellipsoid. The
problem is that we do not know the actual eccentricity and the actual
direction of the rotation axis for any individual galaxy. As long as
this is true, it is in principle impossible to measure the distortion
effect upon any individual galaxy. The only way out of this
difficulty is to use statistical methods in order to measure the
distortion effect upon sufficiently large samples of galaxies. This
approach is based on the assumption that, on a sufficiently large
scale, the rotation axes of galaxies are randomly distributed in the
universe. (This is a reasonable working hypothesis, but it is not
beyond any doubt. In a rotating universe, e.g., the angular momenta
of galaxies are expected to be preferably aligned with the universal
rotation axis.) The distortion pattern, as introduced above, should
then be obtained by averaging galaxy ellipticities over sufficiently
large fields in the sky. In this statistical sense, the distortion
effect has actually been measured by several groups. In the following
paragraphs we give a brief overview of the observational situation.

The first attempt of observing the distortion effect was done already
in 1966 by Kristian \cite{K} with a small number of galaxies in
several clusters. The observation was unsuccessful which, together
with some reasonable assumptions, yielded an upper limit for the
magnetic part of the Weyl tensor. This null result was confirmed in
1983 by Valdes et al. \cite{VALDES} with a 300fold improvement in
accuracy over Kristian, owing to a bigger number of galaxies and
including galaxies at larger distances. The first successful
observation of the distortion effect by statistical methods was
reported in 1990 by Tyson et al. \cite{TYSON} who considered the
distorting effect of rich clusters of galaxies upon background
galaxies. The same mechanism -- background galaxies distorted by the
gravitational field of an intervening cluster -- is usually
considered as an explanation for the socalled {\em giant luminous
arcs\/} which have been discovered in great number since 1986,
beginning with Lynds and Petrosian \cite{LP} and Soucail et al.
\cite{So}. (In the latter case, the linear approximation is, of
course, not applicable, i.e., it is not justified to view all light
rays coming from the source to the observer as `infinitesimally
close' to some central light ray.) In this sense, the distortion
effect produced by galaxy clusters was a well-established phenomenon
by the mid-1990s. On the other hand, by this time all attempts to
observe the distortion effect produced by large-scale structure had
failed, see Mould et al. \cite{MOULD} for an unsuccessful attempt in
1994.

With the recent advancement of CCD mosaic cameras the observational
situation has greatly changed. Now it is
possible to observe `great' parts ($\sim 30\times 30$ square
arc-minutes) of the sky with a large number of galaxies $(\sim
10^5)$ in a single picture and to handle the contained information
automatically. The unwrought data can be cleaned by several
techniques described in \cite{{BACON},{vanW},{KAISER},{WITTMAN}}
which improves the significance of the effect. The results of these
recent observations can be summarized in the following way. A non-zero
distortion effect on a large scale has been measured by demonstrating
a correlation of galaxy ellipticities in several selected fields
in the sky \cite{{SCHNEIDER},{BACON},{vanW},{KAISER},{WITTMAN}},
and there is evidence that this correlation decreases if the field
size is increased \cite{{BACON},{vanW},{KAISER},{WITTMAN}}.
Image distortion by large-scale structure is often called
{\em cosmic shear}, refering to the shear induced in light bundles
on a cosmic scale. This term should {\em not\/} be confused with a
hypothetical shear of the Hubble flow which occasionally is also
called `cosmic shear' or `cosmological shear'.

For future perspectives of observations we refer to the Sloane Digital
Sky Survey (see {\tt http://www.sdss.org}) and to the Deep Lens Survey
(see {\tt http://dls.bell-labs.com}) which are both under way, but also to the
proposed Dark Matter Telescope (see, e.g., Tyson, Wittman and Angel \cite{TWA}).
It does not seem to be overly optimistic to hope that in some years the
observational material on the distortion effect covers large parts of the
sky.

Observations of image distortions are usually evaluated with the help of
the {\em weak-lensing formalism\/} which is based on early ideas of Gunn
\cite{Gun1,Gun2} (also see Webster \cite{WEBSTER}) and was developed
since 1990 by several authors including, e.g., Miralda-Escud\'e
\cite{MIRALDA}, Kaiser \cite{KAI}, van Waerbeke, Bernardeau and
Mellier \cite{BERNARDEAU}; for a comprehensive review we refer to
Bartelmann and Schneider \cite{BS}. This formalism uses Newtonian
approximations in a weakly perturbed Friedmann-Robertson-Walker
model. As the Weyl tensor of a Friedmann-Robertson-Walker model
vanishes, the distortion effect in such a universe is produced by the
perturbations alone. Among other things, the weak-lensing formalism
provides a relation between the correlation function of ellipticities
and the power spectrum of mass fluctuations, thereby relating the
distortion effect to the distribution of (dark) matter. These results
rely, of course, on the Newtonian approximations. The fact that the
distortion effect is determined by the Weyl tensor, on the other
hand, is purely geometric and quite general. In the weak-lensing
formalism this central role of the Weyl tensor is somewhat disguised
and partly overshadowed by the approximations. In particular, the
effect of the magnetic part of the Weyl tensor is completely
neglected because it has no Newtonian counterpart. The present
paper is motivated, at least partly, by our desire to rediscuss
the role of the Weyl tensor for image distortions in view of the
recent observations. More precisely, we want to ask if the observed
image distortion by large-scale structure can be used to gain some
information about the Weyl tensor of the universe. We shall discuss
this question, on the basis of the results derived in the body of
this paper, in the conclusions.

Leaving aside all applications to cosmology, we feel that the
patterns derived and discussed in Section \ref{sec:discussion} have
some general value from a didactical point of view. They illustrate
the image distortion not only in cosmological models but in all spacetimes
of the respective Petrov type (in lowest non-trivial order with respect to the
distance) and the dependence of this effect on the observer's
velocity. This might be helpful for associating the Weyl tensor with
some geometrical imagination.


\section{Derivation of the distortion field}\label{sec:derivation}
    As already mentioned in the introduction, distortion as a
    cosmological observable was introduced in a
    pioneering paper by Kristian and Sachs \cite{KS}. In this section
    it is our goal to rewrite the basic equations by which the
    distortion field is described in terms of Newman-Penrose
    coefficients. In the following section we will then discuss
    these results for each Petrov type. If the reader is not familiar
    with the Newman-Penrose formalism and with the Petrov classification
    he or she may consult Chandrasekhar \cite{Ch}, Chapters 1.8 and 1.9,
    for a detailed introduction. We adopt the same sign and factor
    conventions as Chandrasekhar. In particular, we use the signature
    $(+,-,-,-)$.

    In an arbitrary spacetime (i.e., a 4-dimensional Lorentzian
    manifold), we fix a point $p$ and a pseudo-orthonormal frame
    $(\E_1, \E_2, \E_3, \E_4)$ at $p$. The latter
    may be expressed equivalently in terms of a null tetrad
    $(\el, \n, \m, \q)$ via
\begin{equation}\label{eq:frames}
\begin{split}
    E_1^a = \tfrac{1}{\sqrt{2}} \big( m^a + \bar{m}{}^a \big) \, , \quad
    E_2^a = \tfrac{i}{\sqrt{2}} \big( m^a - \bar{m}{}^a \big) \, ,
\\
    E_3^a = \tfrac{1}{\sqrt{2}} \big( \ell^a - n^a \big) \, , \quad
    E_4^a = \tfrac{1}{\sqrt{2}} \big( \ell^a + n^a \big) \, , \quad
\end{split}
\end{equation}
    cf. Chandrasekhar \cite{Ch}, Chapter 1.8. We interpret the
    timelike vector $\E_4$ as past-pointing so that $-\E_4$ may be
    viewed as the 4-velocity of an observer. Then the totality of
    lightlike geodesics issuing from $p$ into the past is in natural
    one-to-one relation with the set of past-pointing lightlike initial
    vectors
\begin{equation}\label{eq:sphere}
\begin{split}
    k^a = {\mathrm{sin}} \vartheta \big( {\mathrm{cos}} \varphi \, E_1^a
    + {\mathrm{sin}} \varphi \, E_2^a \big) + {\mathrm{cos}} \vartheta \,
    E_3^a + E_4^a \qquad \qquad \quad
\\
    = \tfrac{1}{\sqrt{2}} \big\{ {\mathrm{sin}} \vartheta
    \big( e^{i \varphi} m^a + e^{-i \varphi} {\bar{m}}{}^a \big) +
    \big( {\, 1 \, +\mathrm{cos}} \vartheta  \big) \ell^a +
    \big( \, 1 \, - {\mathrm{cos}} \vartheta \big) n^a \big\} \, ,
\end{split}
\end{equation}
    where $\vartheta$ and $\varphi$ have their usual range as
    standard coordinates on the 2-sphere. We may interpret $\vartheta$
    and $\varphi$ as coordinatizing the {\em celestial sphere\/} of
    the observer with 4-velocity $-\E_4$ at $p$. Every point on the
    past light cone of $p$ can be written as
    ${\mathrm{exp}} (s {\boldsymbol{k}})$ with some parameter $s > 0$
    and some ${\boldsymbol{k}}$ given by (\ref{eq:sphere}); if we restrict
    to sufficiently small $s$, this representation is unique. Henceforth
    we shall use the affine parameter $s$ as a measure for the
    {\em distance\/} from $p$, and it is our goal to discuss the distortion
    effect in lowest non-trivial order with respect to $s$. To within
    this approximation, $s$ can be replaced with the {\em angular
    diameter distance\/} $D = s + O(s^2)$, with the {\em luminosity
    distance\/} $\hat{D} = s + O(s^2)$, or with any other reasonable
    measure of distance used in cosmology. The redshift $z$ is related
    to $s$ by an equation of the form $z = H s + O(s^2)$ with a
    `Hubble constant' $H$ which, in general, is a non-constant function
    on the sky. Also, it is important to realize that $z$ depends not
    only on the velocity of the observer but also on the velocity of the
    light source. (Writing cosmological relations in terms of power series
    was a basic idea introduced by Kristian and Sachs \cite{KS}.
    A discussion of the relations between $s$, $D$, $\hat{D}$ and $z$
    needed here can be found, e.g., in Hasse and Perlick \cite{HP}.)

    At each point of the celestial sphere, the two orthonormal vectors
\begin{equation}\label{eq:tangent}
\begin{split}
    \frac{\partial k^a}{\partial \vartheta} = \tfrac{1}{\sqrt{2}}
    \big\{ {\mathrm{cos}} \vartheta \big( e^{i \varphi} m^a +
    e^{-i \varphi} {\bar{m}}{}^a \big) -
    {\mathrm{sin}} \vartheta \big( \ell^a - n^a \big)\big\} \, ,
\\
    \frac{1}{\mathrm{sin} \vartheta}
    \frac{\partial k^a}{\partial \varphi} = \tfrac{i}{\sqrt{2}}
    \big( e^{i \varphi} m^a - e^{-i \varphi} {\bar{m}}{}^a \big)
    \qquad \qquad
\end{split}
\end{equation}
    span the tangent space to the celestial sphere. (Here and in the
    following it goes without saying that one has to mind the
    familiar coordinate singularities at $\vartheta = 0$ and $\vartheta
    = \pi$.) Hence, every tangent vector to the celestial sphere is of
    the form
\begin{equation}\label{eq:t}
\begin{split}
    t^a = \, u \, \frac{\partial k^a}{\partial \vartheta} \, +
    \, v  \, \frac{1}{\mathrm{sin} \vartheta}
    \frac{\partial k^a}{\partial \varphi}
    \qquad \qquad \qquad \qquad \qquad
\\
    = \tfrac{1}{\sqrt{2}} \big\{
    \big( u \, {\mathrm{cos}} \vartheta + i v \big)
    e^{i \varphi} m^a +
    \big( u \, {\mathrm{cos}} \vartheta - i v \big)
    e^{-i \varphi} {\bar{m}}{}^a -
    u \, {\mathrm{sin}} \vartheta \big( \ell^a - n^a \big) \big\}
\end{split}
\end{equation}
    with $u, v \in {\mathbb{R}}$. Restricting $u$ and $v$ to a circle,
    $u^2 + v^2 \le r^2$, gives an `infinitesimally thin' bundle of light
    rays with an initially circular cross-section. When following those
    rays into the past we will observe, in general, that the cross-section
    does not stay circular but becomes elliptical, owing to the shear
    produced in the bundle by the spacetime geometry. (Please note that,
    according to general relativity, the question of whether an
    `infinitesimally thin' light bundle has a circular shape has an
    observer-independent answer; this was demonstrated already in 1932 by
    Kermack, McCrea and Whittaker \cite{KMW}, cf. Schneider, Ehlers and
    Falco \cite{SEF}, Section 3.4.1.)
    Correspondingly, if the cross-section of the bundle is supposed to be
    circular at some affine distance $s$, then we have to restrict $u$
    and $v$ to a certain ellipse. The equations necessary to calculate
    the shape of this ellipse can be found in the paper by Kristian and
    Sachs \cite{KS}. Their analysis is based on the well-known fact,
    established in a fundamental paper by Sachs \cite{Sa}, that the
    shear of an infinitesimally thin bundle of light rays is governed
    by the Weyl tensor along the central
    ray. More precisely, their result reads as follows. To determine
    an infinitesimally thin bundle which has a circular cross-section
    at some affine parameter distance $s$ from the event $p$, one has
    to consider a certain quadratic form $p_{ab} t^a t^b$ on the
    two-dimensional space of tangent vectors (\ref{eq:t}). In lowest
    non-trivial order with respect to $s$, $p_{ab}$ is given as
\begin{equation}\label{eq:p}
    p_{ab} = - \tfrac{s^2}{6} C_{acbd}k^c k^d + O(s^3) \, ,
\end{equation}
    where $C_{acbd}$ denotes the Weyl tensor at the event $p$, cf.
    eq. (28) in Kristian and Sachs \cite{KS}. If we neglect the
    $O(s^3)$-term, we can read from (\ref{eq:p}) that $p_{ab}$ is
    (real-symmetric and) trace-free; so it has two real eigenvalues
    which differ by sign, say $\frac{s^2}{6} \varepsilon$ and
    $- \frac{s^2}{6} \varepsilon$ with $\varepsilon \ge 0$.
    To get a bundle with circular cross-section at $s$ one has to restrict
    $t^a$ to an ellipse whose major axis is the eigenspace of
    $p_{ab}$ to the eigenvalue $- \frac{s^2}{6} \varepsilon$ and whose
    eccentricity is equal to $\frac{s^2}{6} \varepsilon$.

    The special form of the $O(s^3)$-term in (\ref{eq:p}) will be of
    no interest to us since we will restrict to affine distances which
    are small enough to neglect this term. However, it is interesting
    to note that, if this term is written as a power series in $s$,
    the factor in front of $s^{i+2}$ is a linear function of the
    $i\,$th covariant derivative of the Weyl tensor at $p$. Hence, if the
    Weyl tensor is covariantly constant, then the $O(s^3)$-term in
    (\ref{eq:p}) is zero. Correspondingly, if the Weyl tensor varies
    but little this term may be neglected even for values of $s$ that
    are not small.

\begin{figure}
        \begin{center}
                \setlength{\unitlength}{1.6cm}
                \begin{picture}(6.3,3.2)
                        {\linethickness{0.07cm}
                        \curve(1.54159,0.770796, 4.74159,2.3708)}
                        {\linethickness{0.04cm}
                        \curve(0,0, 0,3.14159)
                        \curve(0,0, 6.28319,0)
                        \curve(6.28319,0, 6.28319, 3.14159)
                        \curve(0,3.14159, 6.28319,3.14159)}
                        {\linethickness{0.02cm}
                        \curve(3.14159,0, 3.14159, 3.14159)
                        \curve(0,1.5708, 6.28319,1.5708)
                        \put(4.2,-0.4){\vector(1,0){1.3}}
                        \put(-0.33,1.17){\vector(0,-1){0.8}}
                        \put(3.46159,1.9308){\vector(2,1){1.22}}
                        \put(2.66159,1.5308){\vector(-2,-1){1.22}}
                        \put(3.14159,1.54159){\arc(1.5,0.02){28}}
                        \put(3.14159, -0.4){$\varphi$}
                        \put(-0.4, 1.5208){$\vartheta$}
                        \put(3.94159, 1.7208){$\chi$}
                        \put(2.8, 1.7){$2 \sqrt{\varepsilon \;}$}}
                \end{picture}
        \end{center}
\caption{ At each point of the celestial sphere, the distortion length
    element is determined by the non-negative number $\varepsilon$
    and the angle $\chi$ (mod $\pi$) }
\label{fig:chi}
\end{figure}
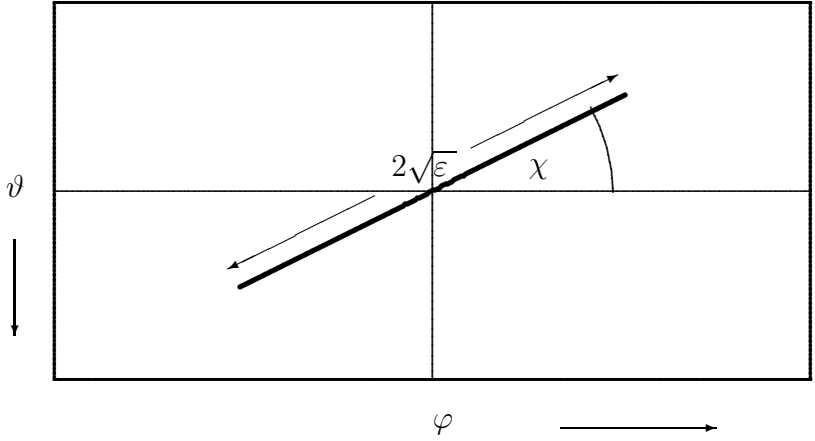

    Now we restrict to light sources at a fixed affine distance
    $s$ and we assume that, for this $s$, the $O(s^3)$-term in
    (\ref{eq:p}) can be neglected. Then the distortion effect is
    completely determined by the Weyl tensor at $p$. Knowing $C_{acbd}$
    allows to calculate the eigenvectors and eigenvalues of (\ref{eq:p}),
    for all values of $\vartheta$ and $\varphi$ (hidden in $k^c k^d$).
    The result may be graphically represented by assigning to each
    point of the celestial sphere a `length element' whose direction
    indicates the eigenspace of $p_{ab}$ to the eigenvalue
    $- \frac{s^2}{6} \varepsilon$ and whose length is a measure
    of $\frac{s^2}{6} \varepsilon$ (for fixed $s$).
    We find it convenient to choose the length elements proportional
    to ${\sqrt{\varepsilon}}$, see Figure \ref{fig:chi}.
    Using the notation of
    this figure, the `distortion length element field' is
    determined by giving the number $\varepsilon$
    and the angle $\chi$ for each point of the celestial sphere.
    Kristian and Sachs \cite{KS} used the quantity $e = 1 + 2 \varepsilon$
    as a measure for the magnitude of the distortion effect. Penrose and Rindler
    \cite{PR}, Volume II, Section 8, discussed the direction field spanned by our
    `distortion length element field', and they called this the {\em fingerprint
    direction field\/} of the Weyl tensor.

    If we exclude the trivial case that the Weyl tensor vanishes at
    $p$, then there are exactly four (not necessarily distinct)
    points on the celestial sphere at which $\varepsilon$ is zero.
    These four points correspond to the four {\em principal null
    directions\/} of the Weyl tensor, and the question of how many
    principal null directions coincide leads to the {\em Petrov
    classification}, see Chandrasekhar \cite{Ch}, Chapter 1.9. In this
    paper it is our main goal to discuss the distortion field over the
    whole sky for the various Petrov types, thereby extending the
    results of Penrose and Rindler \cite{PR}, Volume II, Section 8,
    who analyze the {\em fingerprint direction field\/} near a principal
    null direction. To that end it will be convenient to
    express the distortion field (i.e., $\varepsilon$ and $\chi$ as
    functions of $\vartheta$ and $\varphi$) with the help of the
    Newman-Penrose coefficients
\begin{equation}\label{eq:NP}
\begin{split}
    \Psi_0 = - C_{abcd} \ell^a m^b \ell^c m^d \, , \; \;
    \Psi_1 = - C_{abcd} \ell^a n^b \ell^c m^d \, , \; \;
    \Psi_2 = - C_{abcd} \ell^a m^b \bar{m}{}^c n^d \, ,
\\
    \Psi_3 = - C_{abcd} \ell^a n^b \bar{m}{}^c n^d \, , \; \;
    \Psi_4 = - C_{abcd} n^a \bar{m}{}^b n^c \bar{m}{}^d \, . \;
    \qquad \qquad
\end{split}
\end{equation}
    Inserting (\ref{eq:sphere}) and (\ref{eq:t}) into (\ref{eq:p})
    with the $O(s^3)$-term neglected,
    and using the well-known symmetry properties of the Weyl tensor,
    allows one to express the quadratic form $p_{ab}t^a t^b$ in terms
    of the five complex Newman-Penrose coefficients (\ref{eq:NP}).
    This gives a rather long expression, but with some elementary
    algebra it can be conveniently rewritten as
\begin{equation}\label{eq:pNP}
    p_{ab} t^a t^b = \tfrac{s^2}{6} \, {\mathrm{Re}} \big\{
    (u+iv)^2 \varepsilon e^{-2i \chi} \big\}
\end{equation}
    where $\varepsilon \ge 0$ is the modulus and $- 2 \chi$ (mod $2 \pi$)
    is the argument of the complex number
\begin{equation}\label{eq:general}
\begin{split}
    \varepsilon \, e^{-2i \chi} \; = \;
    \tfrac{1}{2} \, \Psi _0 \,
    (1 + {\mathrm{cos}}\, \vartheta)^2 \, e^{2i \varphi}
    + 2 \, \Psi _1 \, {\mathrm{sin}}\,\vartheta \,
    (1 + {\mathrm{cos}}\, \vartheta) \, e^{i \varphi } + \;
\\
    3 \, \Psi _2 \, {\mathrm{sin}}^2 \vartheta
    + 2 \, \Psi _3 \, {\mathrm{sin}}\,\vartheta \,
    (1 - {\mathrm{cos}}\, \vartheta) \, e^{-i \varphi }
    + \tfrac{1}{2} \, \Psi _4 \,
    (1 - {\mathrm{cos}}\, \vartheta)^2 \, e^{-2i \varphi} \, .
\end{split}
\end{equation}
    Eq. (\ref{eq:pNP}) can be rewritten in matrix notation as
\begin{equation}\label{eq:pmatrix}
    p_{ab} t^a t^b = \tfrac{s^2}{6} \, \varepsilon \;
    ( \, u \; \, v \, )
\begin{pmatrix}
    \; {\mathrm{cos}} \, 2 \chi  & \, {\mathrm{sin}} \, 2 \chi \; \\
        \; {\mathrm{sin}} \, 2 \chi  & -{\mathrm{cos}} \, 2 \chi \;
\end{pmatrix}
\begin{pmatrix}
    u \\ v
\end{pmatrix}.
\end{equation}
    From this equation we can easily deduce that the eigenvalues
    of $p_{ab}$ are $\frac{s^2}{6} \varepsilon$ and
    $- \frac{s^2}{6} \varepsilon$, and that the eigenspace
    to the negative eigenvalue is spanned by
\begin{equation}\label{eq:eigen}
    t^a =
    - {\mathrm{sin}} \chi \, \frac{\partial k^a}{\partial \vartheta}
    + {\mathrm{cos}} \chi \, \frac{1}{\mathrm{sin} \vartheta} \,
    \frac{\partial k^a}{\partial \varphi} \, .
\end{equation}
    This shows that
    $\varepsilon$ and $\chi$, as defined by (\ref{eq:general}), have
    indeed the same meaning as introduced before, see Figure \ref{fig:chi}.
    In other words, if we know the five Newman-Penrose coefficients
    $\Psi_0 , \dots , \Psi_4$ (i.e., if we know the components of
    the Weyl tensor with respect to our null frame), then
    (\ref{eq:general}) gives us the distortion field, i.e.,
    $\varepsilon$ and $\chi$ as functions of $\vartheta$ and
    $\varphi$.

    The distortion field depends, of course, on the frame chosen.
    In terms of the pseudo-orthonormal frame $(\E_1 , \E_2, \E_3 , \E_4)$,
    which is related to the null frame $(\el , \n , \m , \q )$ via
    (\ref{eq:frames}), the behavior of the distortion pattern under
    Lorentz transformations can be summarized in the following way.
    (We are interested only in {\em proper\/} Lorentz transformations,
    i.e., those which do not change the temporal or spatial orientation.)
    Changing the spatial vectors $\E_1,\E_2,\E_3$ and leaving the
    timelike vector $\E_4$ unaffected leads to a rigid rotation of
    the celestial sphere. Changing $\E_4$ leads to a conformal
    transformation of the celestial sphere, see Penrose and Rindler
    \cite{PR}, Volume I, for a detailed discussion, and to a
    corresponding deformation of the distortion pattern. We want to
    discuss this dependence on the observer in some detail for the
    various Petrov types in the next section. As our analysis will
    be based on the representation of the distortion field in
    terms of Newman-Penrose coefficients, we shall need the
    behavior of those coefficients under proper Lorentz transformations.
    The relevant formulae are conveniently listed in the book
    by Chandrasekhar \cite{Ch}, Chapter 1.8, who makes use of
    the fact that all proper Lorentz transformations can be composed
    of (spacetime) {\em rotations of class I, II, and III},
    defined respectively by
\begin{gather}
\label{eq:rI}
    I: \, \el \mapsto \el \, , \;
    \m \mapsto \m + a \el , \;
    \q \mapsto \q + \bar{a} \el \, , \;
    \n \mapsto \n + \bar{a} \m + a \q + |a|^2 \el \, ,
\\
\label{eq:rII}
    II: \, \n \mapsto \n \, , \;
    \m \mapsto \m + b \n \, , \;
    \q \mapsto \q + \bar{b} \n \, , \;
    \el \mapsto \el + \bar{b} \m + b \q + |b|^2 \n \, ,
\\
\label{eq:rIII}
    III: \, \el \mapsto A^{-1} \el \, , \;
    \n \mapsto A \n \, , \;
    \m \mapsto e^{i \Theta} \m \, , \;
    \q \mapsto e^{-i \Theta} \q  \, ,
\end{gather}
    with complex numbers $a$ and $b$ and real numbers $A > 0$
    and $\Theta$ (mod $2 \pi$). The transformation behavior of the
    Newman-Penrose coefficients is for rotations of class I:
\begin{equation}\label{eq:pI}
\begin{split}
    \Psi _0 \mapsto \Psi _0 \, , \quad
    \Psi _1 \mapsto \Psi _1 + {\bar{a}} \Psi _0 \, ,\quad
    \Psi _2 \mapsto
    \Psi _2 + 2 {\bar{a}} \Psi _1 + {\bar{a}}{}^2 \Psi _0 \, ,
    \qquad \qquad
\\
    \Psi _3 \mapsto \Psi _3 + 3 {\bar{a}} \Psi _2 +
    3 {\bar{a}}{}^2 \Psi _1 + {\bar{a}}{}^3 \Psi _0 \, , \quad
    \Psi _4 \mapsto \Psi _4 + 4 {\bar{a}} \Psi _3 +
    6 {\bar{a}}{}^2 \Psi _2 + 4 {\bar{a}}{}^3 \Psi _1 + {\bar{a}}{}^4 \Psi _0 \, ;
\end{split}
\end{equation}
    for rotations of class II:
\begin{equation}\label{eq:pII}
\begin{split}
    \Psi _0 \mapsto \Psi _0 + 4 b \Psi _1 + 6 b^2 \Psi _2 +
    4 b^3 \Psi _3 + b^4 \Psi _4 \, , \quad
    \Psi _1 \mapsto \Psi _1 + 3 b \Psi _2 +
    3 b^2 \Psi _3 + b^3 \Psi _4 \, ,
\\
    \Psi _2 \mapsto \Psi _2 + 2 b \Psi _3 + b^2 \Psi _4 \, , \quad
    \Psi _3 \mapsto \Psi _3 + b \Psi _4 \, , \quad
    \Psi _4 \mapsto \Psi _4 \, ;
    \qquad \qquad \;
\end{split}
\end{equation}
    and for rotations of class III:
\begin{equation}\label{eq:pIII}
\begin{split}
    \Psi _0 \mapsto A^{-2} e^{2 i \Theta} \Psi _0 \, , \quad
    \Psi _1 \mapsto A^{-1} e^{i \Theta} \Psi _1 \, , \quad
    \Psi _2 \mapsto \Psi _2 \, , \quad
\\
    \Psi _3 \mapsto A e^{-i \Theta} \Psi _3 \, , \quad
    \Psi _4 \mapsto A^2 e^{-2 i \Theta} \Psi _4  \, .
    \qquad \qquad \;
\end{split}
\end{equation}


\section{Discussion of the distortion field}\label{sec:discussion}
    Eq. (\ref{eq:general}) gives the distortion field, i.e.,
    $\varepsilon$ and $\chi$ as functions of $\vartheta$ and
    $\varphi$, in an arbitrary frame.
    It is our goal to discuss (\ref{eq:general}) for each Petrov type.
    To that end we first choose a frame such that $\Psi _4 \neq 0$ which
    can always be achieved by a rotation of class I, see (\ref{eq:pI}).
    (Here and in the following we exclude the trivial case where the Weyl
    tensor is zero, i.e., of Petrov type O.) We then apply a rotation of
    class III such that $\Psi _4 = 2$, see (\ref{eq:pIII}). In the
    resulting frame, (\ref{eq:general}) is equivalent to
\begin{equation}\label{eq:roots}
    \varepsilon \, e^{-2i \chi} = \prod_{\mu =1}^4 \;
    \big( z_{\mu} \, \sqrt{1 + {\mathrm{cos}} \, \vartheta \, } \;
    e^{i \varphi /2} -
    \sqrt{1 - {\mathrm{cos}} \, \vartheta \, } \;
    e^{-i \varphi /2} \big)
\end{equation}
    where $z_1,z_2,z_3,z_4$ are the roots of the equation
\begin{equation}\label{eq:z}
    z^4 + 2 \, \Psi _3 \, z^3 + 3 \, \Psi _2 \, z^2
    + 2 \, \Psi _1 \, z + \tfrac{1}{2} \, \Psi _0 \, = \; 0 \;.
\end{equation}
    (To prove this, set the left-hand side of (\ref{eq:z}) equal
    to $(z-z_1)(z-z_2)(z-z_3)(z-z_4)$ and verify that this yields
    the same equations for $z_1,z_2,z_3,z_4$ as equating the
    right-hand sides of (\ref{eq:general}) and (\ref{eq:roots}).)
    From this representation we immediately read that the distortion
    field has exactly four (not necessarily distinct) zeros,
    corresponding to the four principal null directions of the
    Weyl tensor. In particular, we read that at the `north pole'
    $\vartheta = 0$ there is a zero of multiplicity $k$ if and
    only if
\begin{equation}\label{eq:k}
\begin{split}
    \Psi _i = 0 \quad {\mathrm{for}} \; \, i = 0, \dots , k-1 \, ,
\\
    \Psi _k \neq 0 \; . \qquad \qquad
\end{split}
\end{equation}
    Thus, if we know that there is a principal null direction of
    multiplicity $k$, then we know that, in an appropriately
    chosen frame, the distortion pattern is given by
    (\ref{eq:general}) with (\ref{eq:k}) and $\Psi _4 = 2$.
    We shall call such a frame {\em adapted\/} to the
    principal null direction henceforth.

\begin{figure}
\centerline{\epsfig{figure=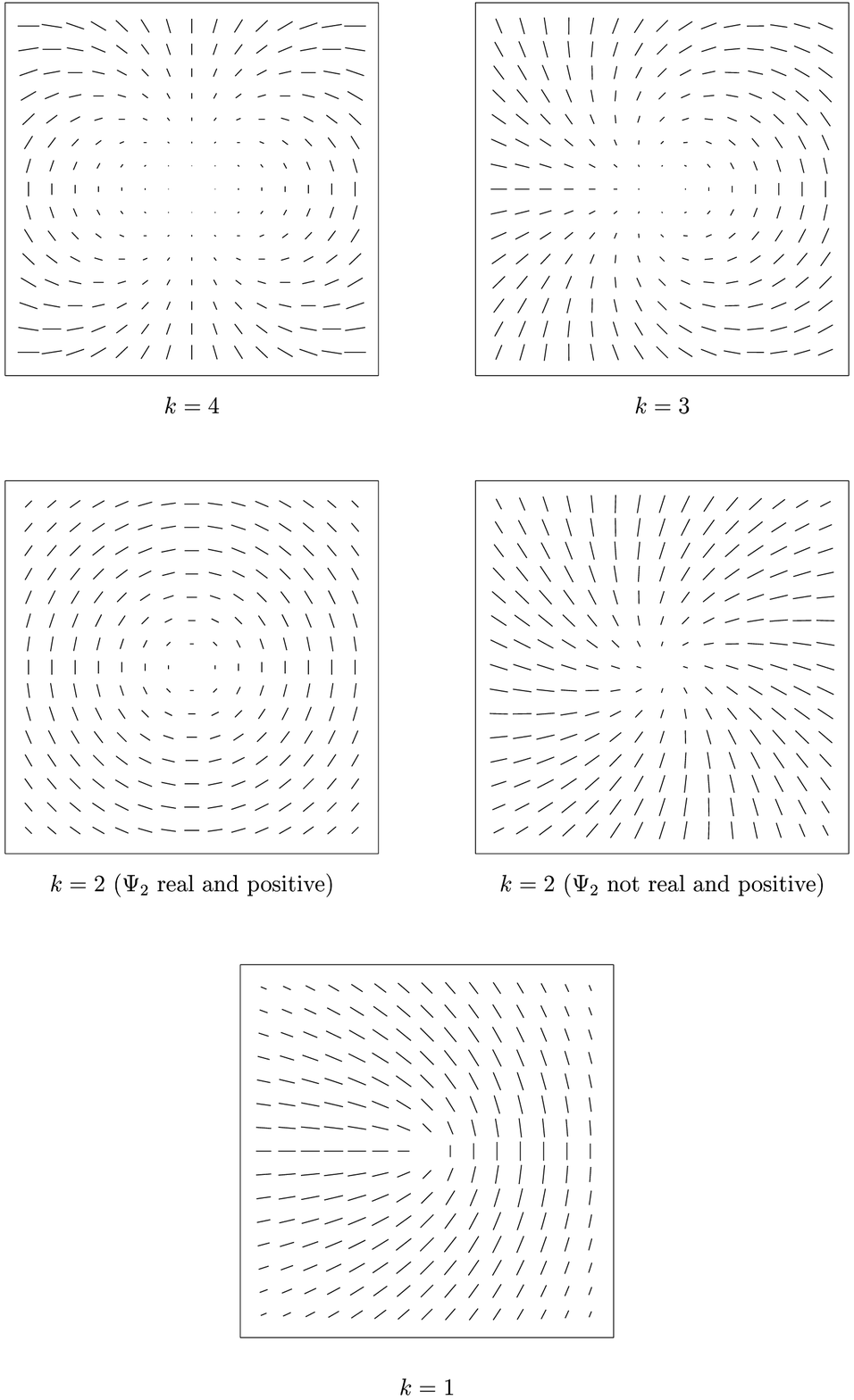, width=16cm}}
\vspace{-2cm}
\caption{ Distortion pattern near a principal null direction of
    multiplicity $k$, cf. Penrose and Rindler \cite{PR}, Volume II,
    Figure~8-3. This picture, like all the following ones, has been
    produced with MATHEMATICA. Choosing in all pictures the length
    elements proportional to ${\sqrt{\varepsilon}}$, rather than to
    $\varepsilon$, was motivated by the fact that otherwise their
    length would vary too strongly, in particular in the case that
    there is a principal null direction of multiplicity 4.}
\label{fig:pol}
\end{figure}

    Before discussing the global features of the distortion patterns
    for each Petrov type, we briefly show that our previous results easily
    allow one to analyze the local structure of the distortion pattern near
    a principal null direction of multiplicity $k$, thereby reproducing
    the results of Penrose and Rindler \cite{PR}, Volume II, Section 8.
    After inserting, for $k=4,3,2,1$, (\ref{eq:k}) and $\Psi _4 = 2$
    into (\ref{eq:general}) we may analyze the resulting pattern
    near $\vartheta = 0$, see Figure~\ref{fig:pol}. In each picture
    $\vartheta$ gives the distance from the center and $\varphi$ is
    the azimuthal angle encircling the center; the length elements
    are proportional to $\sqrt{\varepsilon}$. By inspection we find
    that for $k = 4, 3, 1$ the pattern is uniquely determined, in a
    neighborhood of $\vartheta = 0$, up to diffeomorphisms, i.e.,
    the differential-topological structure of the distortion pattern
    is unique. For $k =2$, the pattern depends on whether or not, in
    the representation of (\ref{eq:k}), the argument of $\Psi _2$
    is zero (mod $2 \pi$), i.e., $\Psi _2$ is real and positive. If
    the argument of $\Psi _2$ is zero (mod $2 \pi$), then the integral
    lines of the distortion field encircle the double zero, otherwise
    they are spiralling into the double zero, with a twist depending on
    the argument of $\Psi _2$. Moreover, from Figure~\ref{fig:pol} we
    read that the distortion field is generated by a continuous vector
    field near a principal null direction of multiplicity $k$ if and only
    if $k$ is even. In other words, only for $k = 4$ and $k =2$ is it
    possible to assign an orientation to the distortion length
    elements in a consistent way. These features have already
    been discussed by Penrose and Rindler. In particular,
    the reader should compare Figure~8-3 in Penrose and Rindler
    \cite{PR}, Volume II, with our Figure~\ref{fig:pol}.

    We shall now discuss the explicit and global structure of the
    distortion pattern for each Petrov type, rather than the
    local differential-topological structure near a principal
    null direction. For each Petrov type, we always start
    out with a frame adapted to a principal null direction of
    highest multiplicity, and we denote the Newman-Penrose coefficients
    in this particular frame by $\hp_0 = 0$, $\hp_1$, $\hp_2$,
    $\hp_3$, and $\hp_4 = 2$. We then apply an arbitrary rotation
    of class I followed by an arbitrary rotation of class III, see
    (\ref{eq:pI}) and (\ref{eq:pIII}). Under such a transformation,
    which is the general form of a proper Lorentz
    transformation keeping the direction of $\el$ fixed, the
    Newman-Penrose coefficients change according to
\begin{gather}
\nonumber
    \Psi _0 = 0 \, , \quad
    \Psi _1 = A^{-1} \, e^{i \Theta} \, \hp _1 \, , \quad
    \Psi _2 = \hp _2 + 2 \, {\bar{a}} \, \hp _1 \, ,
\\
\label{eq:trafo}
    \Psi _3 = A \, e^{-i \Theta} \,
    (\hp _3 + 3 \, {\bar{a}} \, \hp _2 +
    3 \, {\bar{a}}{}^2 \, \hp _1 )  \, ,
\\
\nonumber
    \Psi _4 = A^2 \, e^{-2i \Theta} \,
    (2 + 4 \, {\bar{a}} \hp _3 +
    6 \, {\bar{a}}{}^2 \, \hp _2 +
    4 \, {\bar{a}}{}^3 \, \hp _1 ) \, ,
\end{gather}
    see (\ref{eq:pI}) and (\ref{eq:pIII}). Inserting the new
    Newman-Penrose coefficients into (\ref{eq:general})
    and allowing $A$ to run over ${\mathbb{R}}{}^+$, $\Theta$ over
    $[0, 2 \pi [ \, $, and $a$ over $\mathbb{C}$ will give us the
    distortion pattern in an arbitrary Lorentz frame, with the only
    restriction that we keep the zero of highest multiplicity at the
    north pole. Clearly, this is just a restriction on the choice of
    the spatial axes which is a matter of convenience only. Similarly,
    the effect of the angle $\Theta$ is trivial in the sense that it
    produces just a spatial rotation that can be compensated for by
    a transformation $\varphi \longmapsto \varphi + {\mathrm{const.}}$


\subsection{Type N}\label{subsec:N}
\begin{figure}
\centerline{\epsfig{figure=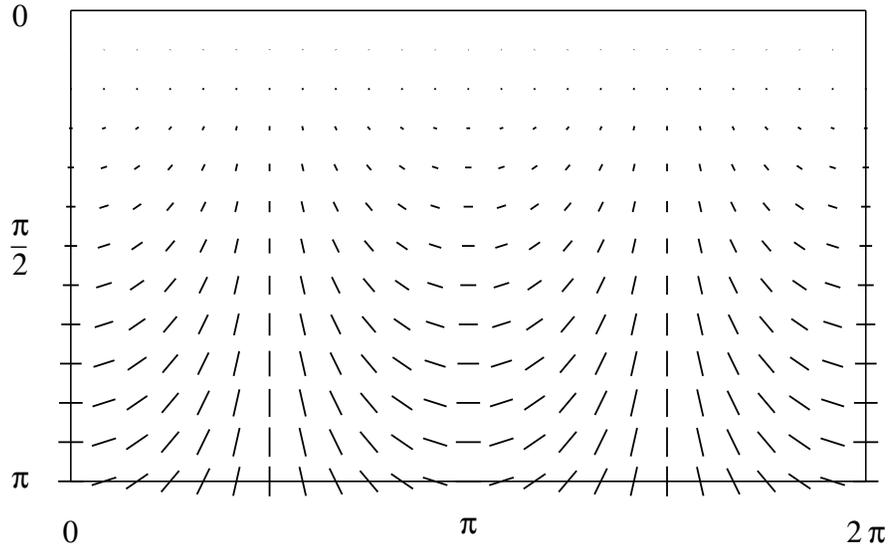, width=12cm}}
\caption{ Distortion pattern for Petrov type N. In this picture, and
    in the following ones, the celestial sphere is given in Mercator
    projection, with $\varphi$ as the horizontal and $\vartheta$ as the
    vertical variable. The picture shows the pattern given by
    (\ref{eq:N}) with $Ae^{- i \Theta} = 0.05$. }
\label{fig:N}
\end{figure}

    Type N is characterized by the fact that all four principal null
    directions coincide. In a frame adapted to this fourfold
    principal null direction we have $\hp_1 = \hp_2 = \hp_3 = 0$ (and,
    of course, $\hp_0 = 0$ and $\hp_4 = 2$). Inserting the
    transformed Newman-Penrose coefficients (\ref{eq:trafo})
    into (\ref{eq:general}) gives us the distortion pattern in
    an arbitrary Lorentz frame (except for spatial rotations),
\begin{equation}\label{eq:N}
    \varepsilon \, e^{-2i \chi} =
    A^2 \, e^{-2i (\Theta + \varphi)} \,
    (1 - {\mathrm{cos}}\, \vartheta)^2 \, ,
\end{equation}
    see Figure~\ref{fig:N}.
    In the hatted frame (i.e., for $A=1$ and $\Theta = 0$),
    (\ref{eq:N}) involves no parameter whatsoever. Hence, the
    distortion pattern for type N is universal in the sense that
    at any two points (in the same or in different spacetimes)
    where the Weyl tensor is of type N we can choose observers
    that see exactly the same distortion patterns. Changing to some
    other observer has the only effect of introducing a scaling factor
    $A$ which is constant over the sky. (The angle $\Theta$ just
    produces a spatial rotation.) By choosing a sequence
    of observers whose 4-velocities approach the fourfold principal
    null direction, the distortion effect can be made arbitrarily
    large, $A \longrightarrow \infty \,$. By choosing a sequence of
    observers whose 4-velocities approach some other null direction,
    the distortion effect can be made arbitrarily small,
    $A \longrightarrow 0 \,$.


\subsection{Type III}\label{subsec:III}
    Type III is characterized by the fact that three principal null
    directions coincide whereas the fourth is different. In a
    frame adapted to the threefold principal null direction we have
    $\hp _1 = \hp _2 = 0$, $\hp _3 \neq 0$ (and $\hp_0 =0$, $\hp_4 = 2$).
    Inserting the transformed Newman-Penrose coefficients
    (\ref{eq:trafo}) into (\ref{eq:general})
    yields
\begin{figure}
\centerline{\epsfig{figure=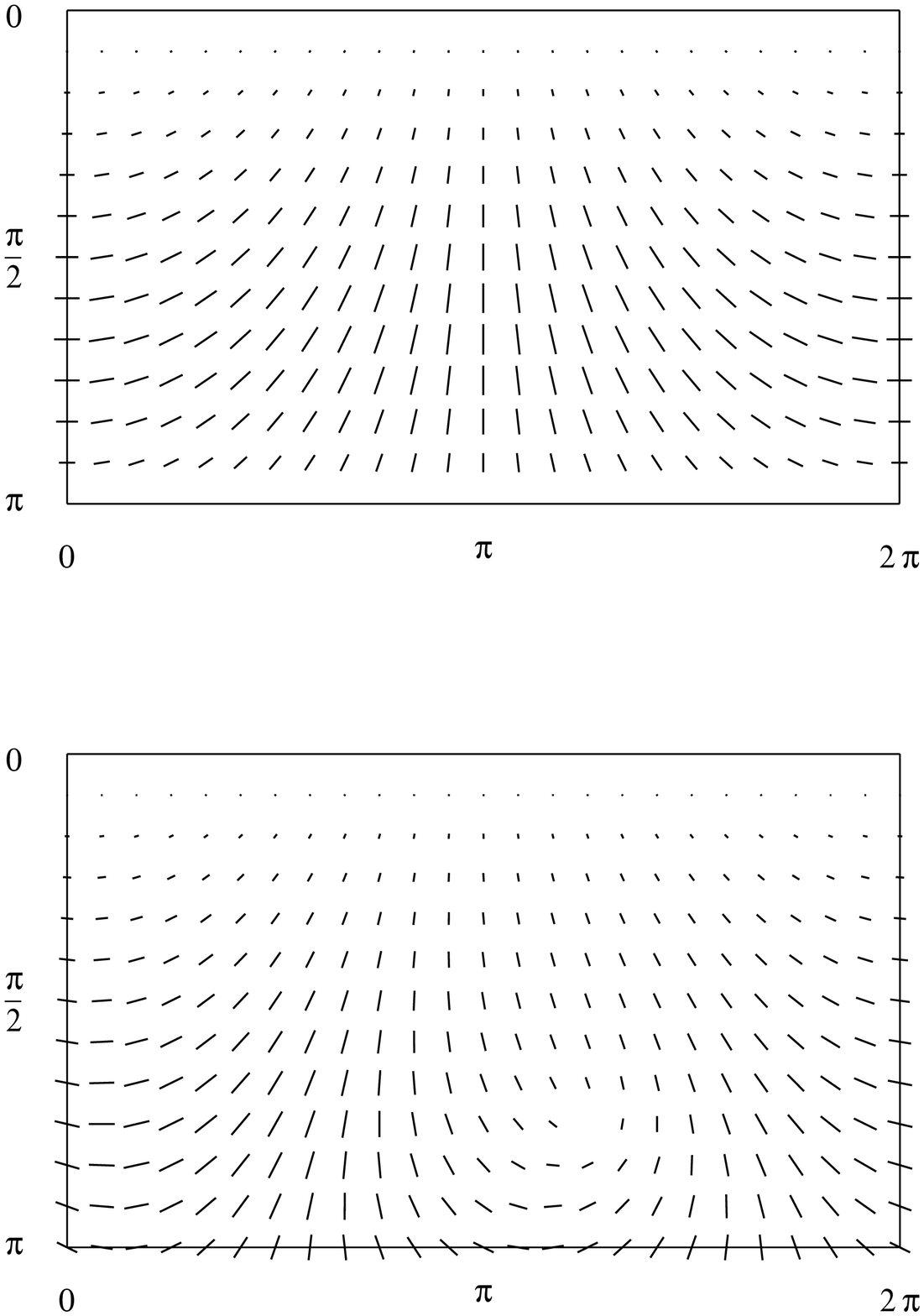, width=14cm}}
\caption{ Distortion pattern for Petrov type III, seen by an observer
    whose 4-velocity lies in the plane spanned by the two different
    principal null directions (top) and by another observer (bottom).
    The first picture shows the pattern $\varepsilon e^{-2 i \chi}
    = 0.007 \; {\mathrm{sin}} \, \vartheta \, ( \, 1 - \, {\mathrm{cos}}
    \, \vartheta \, )$, the second one results by applying a rotation
    of class I with $a = 0.52 - i \, 0.52$, followed by a rotation of
    class III with $A e^{- i \Theta} = 0.56$. Here and in the following
    pictures the apparently strange choice of numerical values is
    motivated by our desire to have the zeros of the distortion pattern
    on (or close to) grid points. }
\label{fig:III}
\end{figure}
\begin{equation}\label{eq:III}
    \varepsilon \, e^{-2i \chi} =
    2 \, A \, e^{-i(\Theta + \varphi )} \, \hp _3 \,
    {\mathrm{sin}}\,\vartheta \, (1 - {\mathrm{cos}}\, \vartheta) +
    A^2 \, e^{-2i (\Theta + \varphi)} \, (1 + 2 {\bar{a}} \hp _3)
    \, (1 - {\mathrm{cos}}\, \vartheta)^2 \, ,
\end{equation}
    see Figure~\ref{fig:III}.
    Whatever the (non-zero) value of $\hp _3$ may be, by choosing $a$,
    $A$ and $\Theta$ appropriately we may give any two values we like
    to the coefficients $A \, e^{- i \Theta} \hp _3 \in {\mathbb{C}}
    \setminus \{ 0 \}$ and $A^2 e^{- 2 i \Theta} (1+2{\bar{a}} \hp _3 )
    \in {\mathbb{C}}$ in (\ref{eq:III}). Thus, type III shows the
    same universality property as type N: For any two points where the
    Weyl tensor is of type III we can find observers that see exactly
    the same distortion patterns. However, the effect of a Lorentz
    transformation is now more complicated than for type N. Choosing
    $a$ such that $1 + \, 2 {\bar{a}} \, \hp _3 = 0$ gives us a frame
    such that the simple zero of the distortion pattern is at the
    south pole ($\vartheta = \pi$), i.e., at the antipodal point of
    the threefold zero. This restriction on the frame, which is
    equivalent to requiring that $\E_4 = \frac{1}{\sqrt{2}} ( \el + \n )$
    lies in the plane spanned by the two different principal null
    directions, still allows to choose $A$ (and $\Theta$) arbitrarily;
    changing $A$ has the same magnifying or demagnifying effect on
    the distortion pattern as for type N (and $\Theta$, as always,
    produces a trivial spatial rotation). If we now choose a different value
    for $a$, thereby boosting the frame such that $\E_4$ is no longer
    in the plane spanned by the two different principal null directions,
    the simple zero of the distortion pattern moves away from the south
    pole. Hence, at a type-III spacetime point the observer can
    easily read from the distortion pattern whether his or her
    4-velocity is in the plane spanned by the two different principal
    null directions.


\subsection{Type D and type II}\label{subsec:DII}
    In a frame adapted to a principal null direction of multiplicity
    two the Newman-Penrose coefficients satisfy $\hp _1 = 0$,
    $\hp _2 \neq 0$ (and $\hp_0=0$, $\hp _4 = 2$). The remaining
    two principal null directions are given by the two non-zero
    solutions of equation (\ref{eq:z}) with the hatted coefficients,
    i.e.,
\begin{equation}\label{eq:z43}
    z_{3/4} = - \hp _3 \pm \sqrt{\hp _3^2 - 3 \, \hp _2 \, } \; .
\end{equation}
    If these two solutions coincide, the Weyl tensor is of type D,
    otherwise it is of type II. Hence, we have
\begin{gather}
\label{eq:rootsD}
    {\mathrm{Type \; D \, :}} \qquad \; 3 \, \hp _2 = \hp _3 ^2
\\
\label{eq:rootsII}
    {\mathrm{Type \; II \, :}} \qquad \, 3 \, \hp _2 \neq \hp _3 ^2.
\end{gather}
    Inserting the transformed Newman-Penrose coefficients
    (\ref{eq:trafo}) into (\ref{eq:general})
    yields the distortion pattern for type D and type II in an
    arbitrary frame,
\begin{equation}\label{eq:II}
\begin{split}
    \varepsilon \, e^{-2i \chi} \; = \;
    3 \, \hp _2 \, {\mathrm{sin}}^2 \vartheta
    + 2 \, A \, e^{-i(\Theta + \varphi)} \,
    (\hp _3 + 3 \, {\bar{a}} \, \hp _2 ) \,
    {\mathrm{sin}}\,\vartheta \, (1 - {\mathrm{cos}}\, \vartheta) +
\\
    A^2 \, e^{-2i(\Theta + \varphi )} \,
    (1 + 2 \, {\bar{a}} \, \hp _3 + 3 \, {\bar{a}}{}^2 \, \hp _2 \, ) \,
    (1 - {\mathrm{cos}}\, \vartheta)^2 \, . \qquad \qquad \quad
\end{split}
\end{equation}
    We first discuss type D. With (\ref{eq:rootsD}), (\ref{eq:II})
    simplifies to
\begin{figure}
\centerline{\epsfig{figure=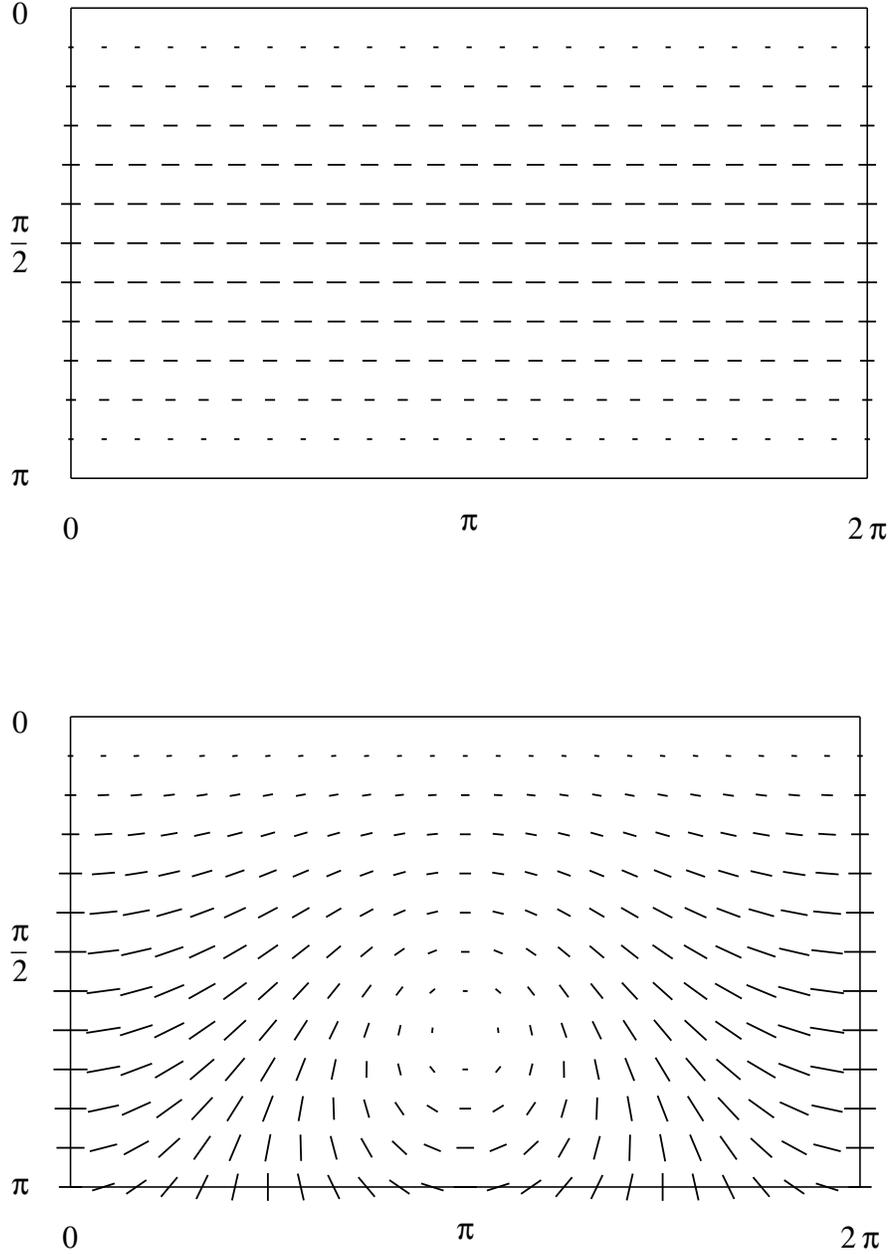, width=14cm}}
\caption{ Distortion pattern for Petrov type D with $\hp _2 =
    \tfrac{1}{3} \hp _3 ^2$ real and positive, seen by an observer
    whose 4-velocity lies in the plane spanned by the two different
    principal null directions (top) and by another observer (bottom).
    We have chosen $\hp _3 = 0.08$. The second picture results from
    the first one by applying a rotation of class I followed by a
    rotation of class III with ${\bar{a}} A e^{- i \Theta} = 0.58$. }
\label{fig:Da}
\end{figure}
\begin{figure}
\centerline{\epsfig{figure=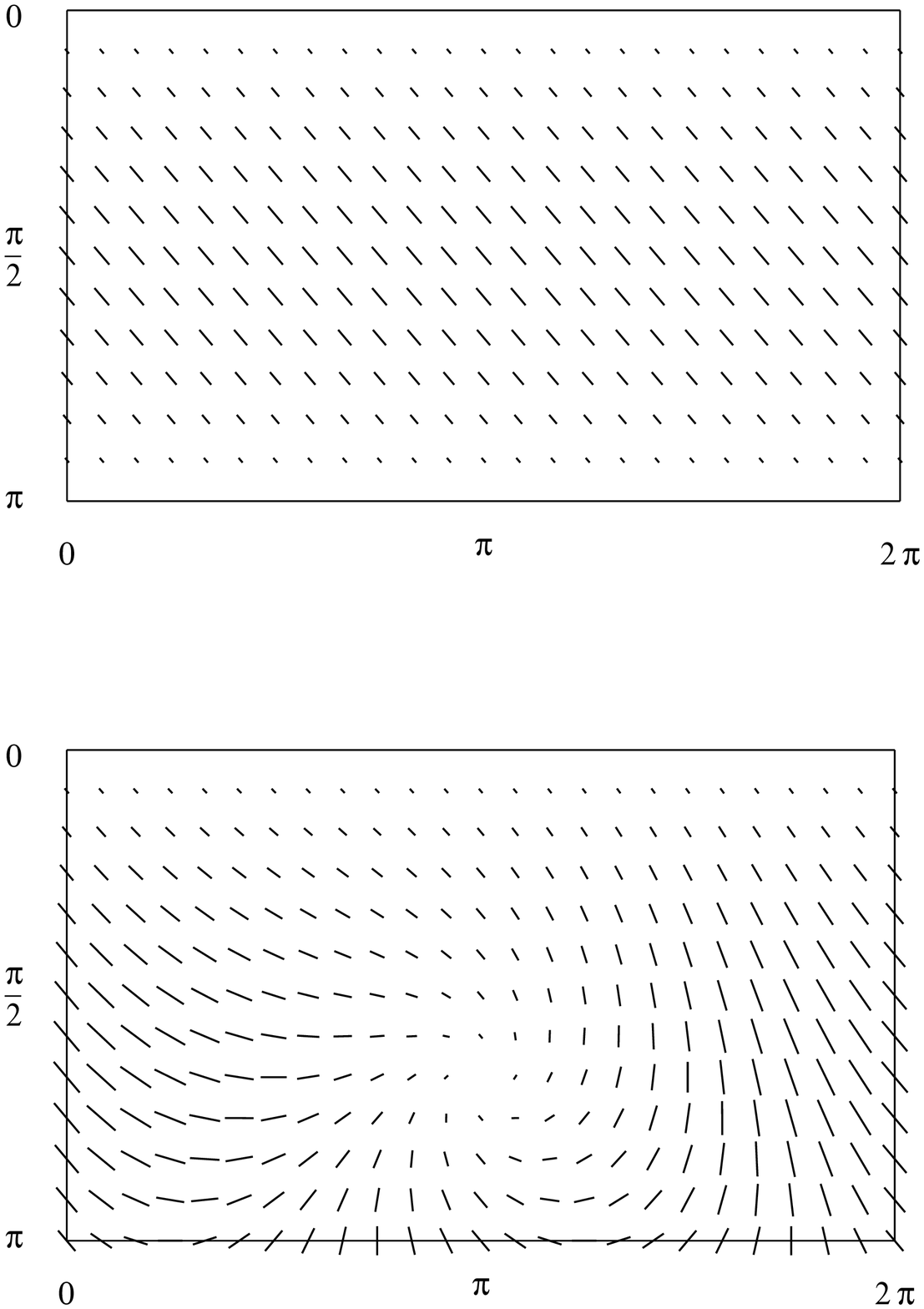, width=14cm}}
\caption{ Distortion pattern for Petrov type D with $\hp _2 =
    \tfrac{1}{3} \hp _3 ^2$ not real and positive, seen by an observer
    whose 4-velocity lies in the plane spanned by the two different
    principal null directions (top) and by another observer (bottom).
    This time we have chosen $\hp _3 = 0.08 e^{i \pi /4}$. As in Figure
    \ref{fig:Da}, the second picture results from
    the first one by applying a rotation of class I followed by a
    rotation of class III with ${\bar{a}} A e^{- i \Theta} = 0.58$. }
\label{fig:Db}
\end{figure}
\begin{figure}
\centerline{\epsfig{figure=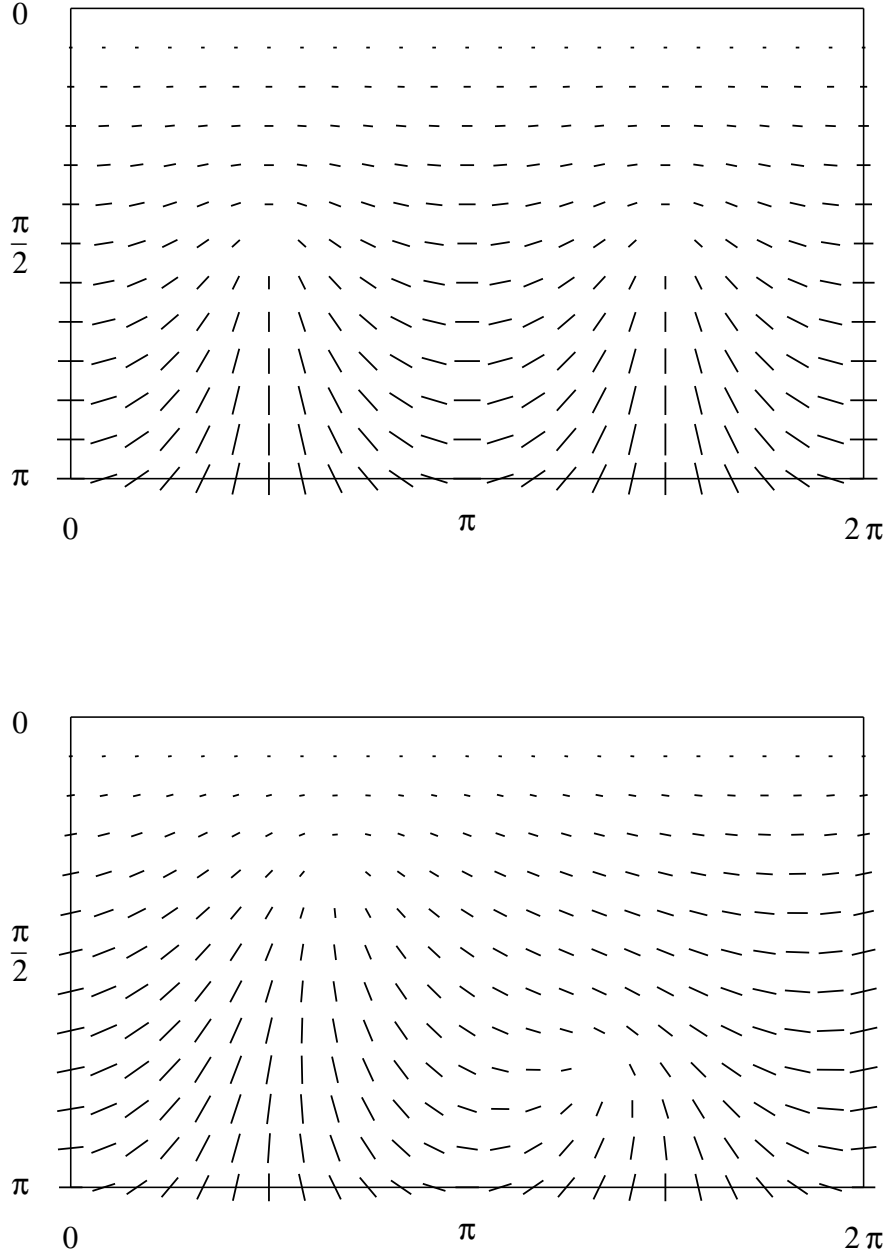, width=14cm}}
\caption{ Distortion pattern for Petrov type II with $\hp _2$ real
    and positive, seen by an observer with the special 4-velocity
    such that the distortion pattern is given by (\ref{eq:IIs})
    (top) and by another observer (bottom). We have chosen $\hp _2
    = 0.001$. $\hp _3$ is arbitrary except for $3 \hp _2 \neq
    \hp _3^2$. The second picture results from the first one
    by applying a rotation of class I with $a = 0.32 + i \, 0.73$,
    followed by a rotation of class III with $A e^{-i \Theta} =
    0.93 + i \, 0.33$. }
\label{fig:IIa}
\end{figure}
\begin{figure}
\centerline{\epsfig{figure=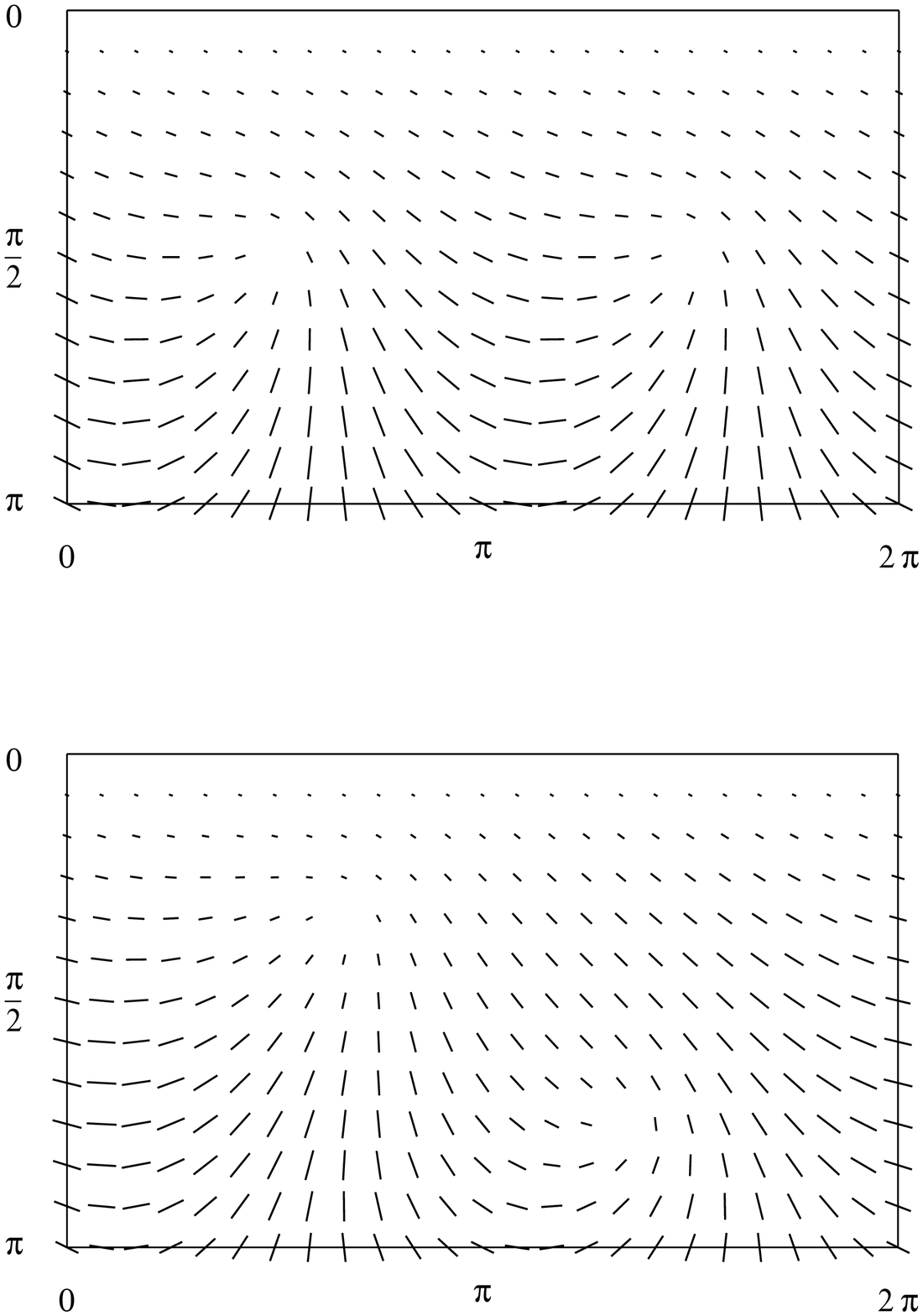, width=14cm}}
\caption{ Distortion pattern for Petrov type II with $\hp _2$ not real
    and positive, seen by an observer with the special 4-velocity
    such that the distortion pattern is given by (\ref{eq:IIs})
    (top) and by another observer (bottom). This time we have
    chosen $\hp _2 = 0.001 e^{i \pi /4}$. Again, $\hp _3$ is arbitrary,
    except for $3 \hp _2 \neq \hp _3^2$, and the second picture follows
    from the first one by applying a rotation of class I with
    $a = 0.32 + i \, 0.73$, followed by a rotation of class III with
    $A e^{-i \Theta} = 0.93 + i \, 0.33$. }
\label{fig:IIb}
\end{figure}
\begin{equation}\label{eq:D}
    \varepsilon \, e^{-2i \chi} =
    \big( \hp _3 \, {\mathrm{sin}} \, \vartheta
    + A \, e^{-i(\Theta + \varphi)} \, (1 + {\bar{a}} \, \hp _3 )
    \, (1 - {\mathrm{cos}}\, \vartheta) \big) ^2,
\end{equation}
    see Figure~\ref{fig:Da} and Figure~\ref{fig:Db}.
    From this equation we read that for type D the universality
    property known from types N and III is not satisfied: The
    distortion pattern depends on the (non-zero) value of $\hp _3$,
    i.e., there is a one-complex-parameter family of genuinely distinct
    type-D spacetime points. Even the differential-topological structure
    of the distortion pattern depends on $\hp _3$. If $\hp _3$ is
    real (i.e., if $\hp _2 = \tfrac{1}{3} \hp _3 ^2$ is real and
    positive), then the integral lines of the distortion field are
    closed curves encircling each of the two double zeros; if $\hp _3$
    is non-real, those integral lines start at one double zero and
    terminate at the other. The effect of a Lorentz transformation
    is as follows. Choosing $a$ such that $1 + {\bar{a}} \, \hp _3
    = 0$ gives us a frame such that the two double zeros are in
    antipodal positions at the sky. This is the case if and only if
    the vector $\E_4 = \frac{1}{\sqrt{2}} ( \el + \n )$ lies in the plane
    spanned by the two double principal null directions. Restricting
    the frames in this way leaves the freedom of choosing $A$ and $\Theta$
    arbitrarily; this, however, has no effect on the distortion pattern.
    In other words, all observers whose 4-velocity is in the plane
    spanned by the two double principal null directions see exactly
    the same pattern. Changing $a$, i.e., boosting to a frame such
    that $\E_4$ is not in the plane spanned by the two double principal
    null directions, has the effect that the second double zero moves
    away from the south pole. Hence, by observing the distortion
    pattern an observer can find out whether the Weyl tensor at
    the point of observation is of type D, and whether
    his or her 4-velocity is in the plane spanned by the two double
    principal null directions. Moreover, the differential-topological
    structure of the pattern alone allows to find out whether in a
    frame adapted to one of the two double principal null directions
    the Newman-Penrose coefficient $\hp _2 = \tfrac{1}{3} \hp _3 ^2$
    is real and positive.

    We now turn to type II, see Figure~\ref{fig:IIa} and
    Figure~\ref{fig:IIb}. By choosing $a$ such that $\hp _3 +
    3 \, {\bar{a}} \, \hp _2 = 0$ and choosing $A$ and $\Theta$ such
    that $A^2 \, e^{-2i \Theta} \, ( 3 \, \hp _2 - \hp _3^2) =
    9 \hp ^2_2$ we may boost to a frame such that
    $\E _4 = \frac{1}{\sqrt{2}} ( \el + \n )$ lies in the plane
    ${\mathcal{H}}$ spanned by the two simple principal null
    directions and the double principal null direction
    is spanned by the sum of $\E _4$ and a vector perpendicular to
    ${\mathcal{H}}$. These properties fix $\E _4$ uniquely. In this
    frame, the distortion field is given by
\begin{equation}\label{eq:IIs}
    \varepsilon \, e^{-2i \chi} =
    3 \, \hp _2 \, \big( {\mathrm{sin}}^2 \vartheta
    +  e^{-2i \varphi } \,
    (1 - {\mathrm{cos}}\, \vartheta)^2 \big) \, .
\end{equation}
    In this representation the two simple principal null directions are
    situated at the equator $\vartheta = \pi /2$ and opposite to each
    other. Similarly to the type-D case, there is a one-complex-parameter
    family of genuinely distinct type-II spacetime points, depending
    on the value of $\hp _2$. Again, the differential-topological
    structure of the distortion pattern depends on whether or not
    $\hp_2$ is real and positive. If this is true, then all the integral
    lines of the distortion field are closed curves encircling the double
    zero, with the exception of one particular integral line that
    connects the two simple zeros. If $\hp _2$ is not real and positive,
    then every integral line connects the double zero with one of the
    two simple zeros.


\subsection{Type I}\label{subsec:I}
    Type I is characterized by the fact that all four principal null
    directions are distinct. In a frame adapted to one of them we have
    $\hp_1 \neq 0$ (and $\hp_0=0$, $\hp_4 =2$). With the transformed
    Newman-Penrose coefficients (\ref{eq:trafo}) we find the following
    expression for the distortion pattern
\begin{figure}
\centerline{\epsfig{figure=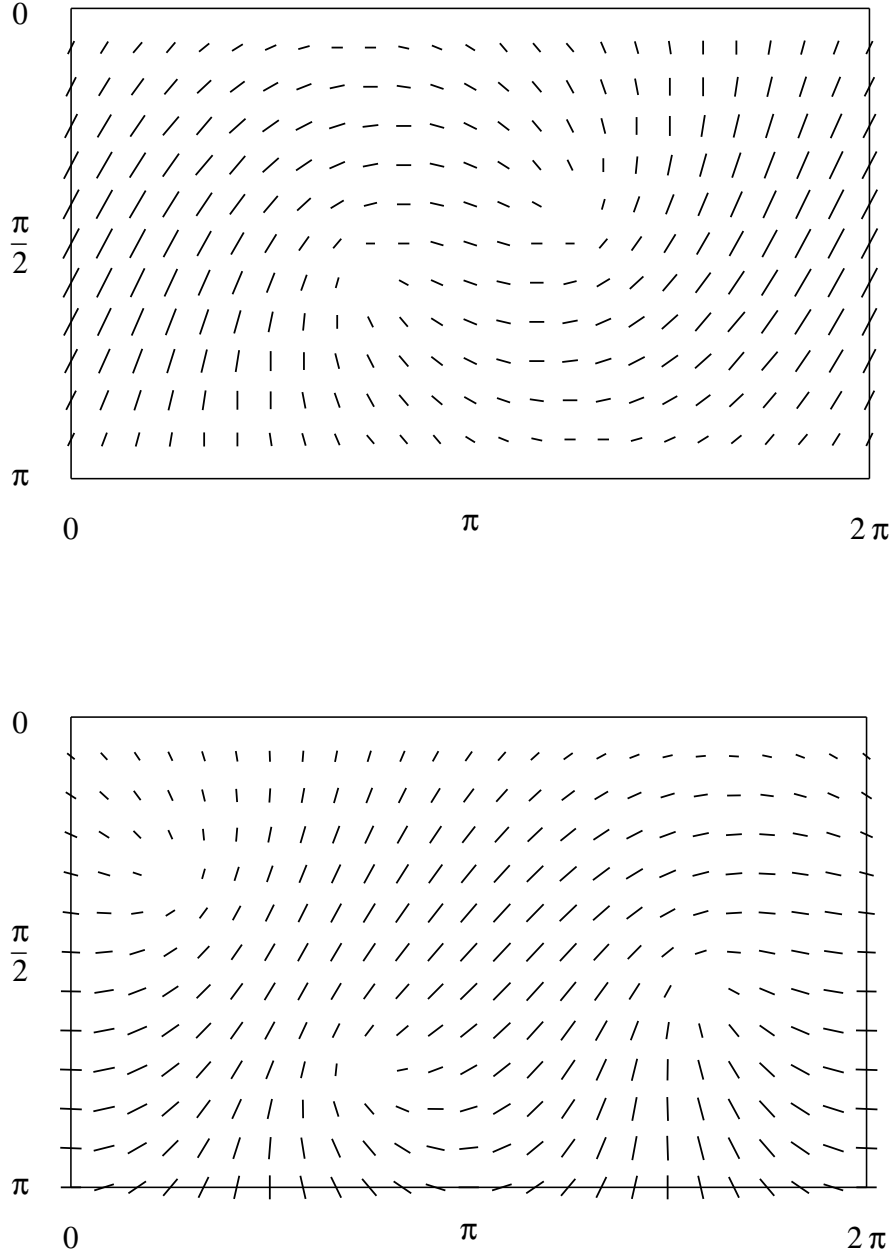, width=14cm}}
\caption{ Distortion pattern for Petrov type I, seen by an observer with
    the special 4-velocity such that the distortion pattern is given
    by (\ref{eq:Is}) (top) and by another observer (bottom). We have
    chosen $P = - 0.0018 - i \, 0.0056$ and $Q = - 0.0022 - i \, 0.0035$.
    The second picture results from the first one by applying a rotation
    of class I with $a = - 0.25 + i \, 0.14$, followed by a rotation
    of class III with $A e^{- i \Theta } = - 1.37 + i \, 0.41$. }
\label{fig:I}
\end{figure}
\begin{gather}
\nonumber
    \varepsilon \, e^{-2i \chi} \; = \;
    2 \, A^{-1} \, e^{i(\Theta + \varphi )} \, \hp _1 \,
    {\mathrm{sin}}\,\vartheta \, (1 + {\mathrm{cos}}\, \vartheta)
    + 3 ( \hp _2 + 2 \, {\bar{a}} \, \hp _1 ) \,
    {\mathrm{sin}}^2 \vartheta +
\\
\label{eq:I}
    2 \, A \, e^{-i( \Theta + \varphi )} \,
    ( \hp _3 + 3 \, {\bar{a}} \, \hp _2 +
    3 \, {\bar{a}}{}^2 \, \hp _1 ) \,
    {\mathrm{sin}}\,\vartheta \, (1 - {\mathrm{cos}}\, \vartheta) +
\\
\nonumber
    A^2 \, e^{-2i ( \Theta + \varphi )} \,
    ( 1 + 2 \, {\bar{a}} \, \hp _3 + 3 \, {\bar{a}}{}^2 \, \hp _2
    + 2 \, {\bar{a}}{}^3 \, \hp _1 )
    (1 - {\mathrm{cos}}\, \vartheta)^2 \, .
\end{gather}
    We may choose $a$ such that $1 + 2 \, {\bar{a}} \, \hp _3 +
    3 \, {\bar{a}}{}^2 \, \hp _2 + 2 \, {\bar{a}}{}^3 \, \hp _1 = 0$,
    thereby moving the second zero to the south pole $\vartheta = \pi$.
    With $a$ fixed that way, we may choose $A$ and $\Theta$ such
    that $A^{-2} \, e^{2i \Theta }\hp_1 =
    ( \hp _3 + 3 \, {\bar{a}} \, \hp _2 + 3 \, {\bar{a}}{}^2 \,\hp _1 )$,
    thereby moving the remaining two simple zeros in symmetric
    positions, $\vartheta _3 + \vartheta _4 = \pi$ and $\varphi _3 +
    \varphi _4 = 2 \pi$. The resulting frame is characterized by the fact
    that $\E _4 = \frac{1}{\sqrt{2}} ( \el + \n )$ lies in the plane
    ${\mathcal{H}}$ spanned by two of the simple principal null
    directions and the other two simple principal null
    directions are spanned by vectors of the form $\E _4 + \X + \Y$
    and $\E _4 + \X - \Y$ where $\X$ and $\Y$ are perpendicular to
    $\E _4$ and $\X$ is perpendicular to ${\mathcal{H}}$. These
    properties characterize $\E _4$ uniquely, except for the possibility
    of permutating the four principal null directions. In this frame,
    the distortion pattern takes the form
\begin{equation}\label{eq:Is}
    \varepsilon \, e^{-2i \chi} =
    Q \, {\mathrm{sin}}\,\vartheta \,
    \big( (1 + {\mathrm{cos}}\, \vartheta) \, e^{i \varphi } +
    (1 - {\mathrm{cos}}\, \vartheta) \, e^{-i \varphi } \big)
    + P \, {\mathrm{sin}}^2 \vartheta
\end{equation}
    where $Q = 2 \, A^{-1} e^{i \Theta} \, \hp _1 \, \in \, {\mathbb{C}}
    \setminus \{ 0 \}$ and $P = 3 \, ( \, \hp _2 \, + \, 2 \,
    \bar{a} \, \hp _1 \, ) \, \in \, {\mathbb{C}}$
    are parameters that characterize the spacetime point. Thus, there
    is a two-complex-parameter family of type-I spacetime points which
    are genuinely distinct in the sense that their distortion patterns
    are not related by a Lorentz transformation (i.e., by a conformal
    transformation of the celestial sphere). However, the distortion
    patterns of type-I spacetime points are always related by
    diffeomorphism, i.e., the differential-topological properties
    of the distortion pattern are the same for all values of $Q$ and
    $P$. The top part of Figure~\ref{fig:I} shows the distortion pattern
    given by (\ref{eq:Is}) for some values of $Q$ and $P$; the bottom
    part demonstrates the effect of a Lorentz transformation on this
    pattern. Two real parameters are necessary for fixing the positions
    of the zeros in the first picture. The other two real parameters
    in $Q$ and $P$ fix the overall scaling factor and the twist of the
    pattern.

    As an aside, we mention that there is always a frame in which
    the four simple zeros on the sky form a {\em disphenoid}, i.e.,
    a tetrahedron such that opposite edges have equal length, see
    Penrose and Rindler \cite{PR}, Volume II, Section 8. However,
    with our convention of keeping one of the zeros at the north
    pole the distortion field in this particular frame is given
    by a rather awkward expression.


\section{Conclusions}\label{sec:conclusions}

    In this paper we have given a new formula, eq. (\ref{eq:general}),
    for the distortion effect on light sources around an observer in an
    arbitrary spacetime. This formula, which is exact to within lowest
    non-trivial order with respect to the distance between light sources
    and observer, comes about by rewriting classical results of Kristian
    and Sachs \cite{KS} in terms of the Newman-Penrose formalism. On the basis
    of this formula we have discussed the distortion patterns graphically
    for each Petrov type. We feel that the resulting pictures have some
    didactic value as visualizing the effect of the Weyl tensor on light
    rays. In this respect our results extend the work by Penrose and
    Rindler~\cite{PR} on the `fingerprint direction field' of the
    Weyl tensor insofar as (i) we discuss not only the direction but
    also the magnitude of the distortion, (ii) our patterns are
    metrically correct on the whole celestial sphere, not just
    differential-topologically near a principal null direction, (iii)
    we discuss the dependence on the observer's velocity.

    As a special application, it is an interesting question to ask whether
    the distortion patterns derived in this paper have some relevance
    in view of cosmology. As mentioned already in the introduction,
    image distortion produced by large-scale structure has been observed
    recently \cite{{SCHNEIDER},{BACON},{vanW},{KAISER},{WITTMAN}} by
    statistically evaluating galaxy ellipticities in selected fields on
    the sky. Following the original ideas of Kristian and Sachs \cite{KS},
    we want to ask whether observations of this kind can be used
    to determine the Weyl tensor `here and now' in an appropriately
    smoothed universe. For discussing this question we subdivide our
    celestial sphere into fields of some specified size and we specify
    a maximal distance up to which galaxies are to be observed. By
    averaging galaxy ellipticities over each field and assigning the
    result to the field center we get the distortion pattern of a
    smoothed universe. Here the smoothing procedure refers to averaging
    the Weyl tensor over conic regions, with the observer at the tip
    of the cone, which is not directly related to averaging the matter
    density. If the field size and/or the maximal distance has been
    chosen small, then the distortion effect in each conic region is essentially
    determined by local inhomogeneities. In other words, even in the smoothed
    universe the Weyl tensor will have a fairly strong variation, so the
    $O(s^3)$-terms in (\ref{eq:p}), which involve covariant derivatives of
    the Weyl tensor, cannot be neglected. For this reason we cannot expect
    that the resulting pattern coincides with one of the patterns derived
    in the preceding section. However, it does make sense to compare the
    observed distortion pattern with the ones derived in the preceding
    section if we increase the field size and the maximal distance
    sufficiently such that the distortion effect of local inhomogeneities
    becomes irrelevant. There are three possible outcomes.

    First, the distortion effect may vanish over the whole sky, for field
    size and maximal distance chosen appropriately. This would indicate that,
    at this level of averaging, the Weyl tensor is zero. This is what we
    expect if we rely on the assumption that our universe differs from a
    Robertson-Walker cosmos only by local `lumps' whose effect on light
    rays averages to zero. The present observations are in accordance with
    this assumption, but one has to keep in mind that they cover only a
    very small portion of the sky. When supported by future observations
    that cover large parts of the sky, this can be viewed as an independent
    confirmation of the standard cosmology assumptions. It is remarkable
    that this line of thought is based on geometry only, i.e.,
    Einstein's field equation is not used.

    Second, the distortion pattern may become similar to one of the patterns
    derived in the preceding section, for some field size and some maximal
    distance. This would indicate that in our standard cosmology model the
    Robertson-Walker background has to be replaced by some other background
    model, namely by one with a non-trivial Petrov type. In this case we
    could read from the distortion pattern not only the Petrov type of the
    Weyl tensor `here and now' (in the smoothed universe) but also the 
    position of our 4-velocity vector in relation to the principal null 
    directions.

    Third, the distortion effect may be non-zero but different from all
    the patterns derived in the preceding section, for any field size and
    any maximal distance. In this case there would be the following
    possible explanations. (i) The galaxies inside one field have non-randomly
    distributed actual axes, even for large field size and large maximal
    distance. This could be taken as indicating a prefered direction in the
    universe such as given, e.g., by a cosmic rotation or by a cosmic magnetic
    field. (ii) Even in the smoothed universe the derivative terms of the
    Weyl tensor are so large that the $O(s^3)$ terms in (\ref{eq:p}) cannot
    be neglected. (iii) In some fields the average distance of the galaxies
    is larger than in other fields. As long as the $O(s^3)$ terms can be
    neglected, this has an effect only on the magnitude but not on the direction
    of the distortion effect, i.e., it has an effect on $\varepsilon$, as a
    function on the celestial sphere, but not on $\chi$.

    In this sense, observing the distortion effect can provide us with some
    information on the Weyl tensor in the universe, based on geometry
    arguments alone. This line of thought is meant as complementary to
    the ongoing efforts of using distortion observations for tracing the
    matter distribution in the universe, based on the approximative assumptions
    of the weak-lensing formalism.

    Two questions are left open for future work. The first one is whether
    there are some classes of spacetimes, e.g. Bianchi models, for which the
    distortion patterns can be calculated {\em exactly}, not just to within lowest
    order with respect to the distance. The second one is whether distortion patterns
    can be linked to the matter distribution without making approximative
    assumptions of the weak-lensing kind. It is true that, quite generally,
    the Weyl tensor is that part of the curvature which is {\em not\/}
    determined, at a particular event, by the energy-momentum tensor at
    that event via Einstein's field equation. However, the divergence of the
    Weyl tensor is related to derivatives of the energy-momentum tensor by
    the equation $C^{abcd}{}_{;d}=T^{c[a;b]}- \frac{1}{3}g^{c[a}T^{;b]}$ which
    follows from the Bianchi identity and Einstein's field equation (see Kundt
    and Tr\" umper \cite{KUNDT} or Ellis \cite{ELLIS}). Maybe this equation
    can be used, at least in special classes of spacetimes, for gaining some
    information about the energy-momentum tensor from distortion patterns. In
    this connection recent work by Frittelli, Kling and Newman \cite{FKN1, FKN2}
    is of some interest. In these articles the authors discuss image distortion
    in arbitrary spacetimes without using approximative assumptions of the
    weak-lensing kind and also without using series expansions of the
    Kristian-Sachs type.

    \subsection*{Acknowledgment}
    T. C. would like to thank the Deutsche Forschungsgemeinschaft for
    supporting this research with the grant HE 1922/5-1.

\end{document}